\documentclass[11pt]{article}
\usepackage{amsmath}
\usepackage{graphicx}
\usepackage{amssymb}
\def\lsim{\mathrel{\raise.3ex\hbox{$<$\kern-.75em\lower1ex\hbox{$\sim$}}}}
\def\gsim{\mathrel{\raise.3ex\hbox{$>$\kern-.75em\lower1ex\hbox{$\sim$}}}}

\newcommand{\be}{\begin{equation}}
\newcommand{\ee}{\end{equation}}
\newcommand{\bea}{\begin{eqnarray}}
\newcommand{\eea}{\end{eqnarray}}
\begin{document}

\title
{Eikonal contributions to ultra high energy neutrino-nucleon cross
sections in low scale gravity models}
\author{ E. M. Sessolo and D. W. McKay\\[2ex]
\small\it Department of Physics and Astronomy, University of
Kansas, Lawrence, KS 66045}

\date{}

\maketitle

\begin{abstract}

We calculate low scale gravity effects on the cross section for
neutrino-nucleon scattering at center of mass energies up to the
Greisen-Zatsepin-Kuzmin (GZK) scale, in the eikonal approximation.
We compare the cases of an infinitely thin brane embedded in $n=5$
compactified extra-dimensions, and of a brane with a physical
tension $M_{S}=1$ TeV and $M_{S}=10$ TeV. The extra dimensional
Planck scale $M_{D}$ is set at $10^{3}$ GeV and $2\times10^{3}$
GeV. We also compare our calculations with neutral current
standard model calculations in the same energy range, and compare the
thin brane eikonal cross section to its saddle point approximation. New
physics effects enhance the cross section by orders of magnitude on
average. They are quite sensitive to $M_{S}$ and $M_{D}$ choices,
though much less sensitive to $n$.

\end{abstract}

\section{Introduction}
Since neutrinos interact only weakly with matter, neutrino
observatories provide a powerful tool for exploring the deepest
reaches of stars and galaxies. In recent years there has been an
upswell in the number of experiments, either running or about to
run, that are aimed at detecting ultra-high energy (UHE)
neutrinos, neutrinos with energies from the multi-TeV range to
beyond the EeV range. Their existence is predicted by a number of
theoretical models. Although the production mechanism can be
different - either as a byproduct of creation and decay of pions
in the interaction between high energy primary cosmic rays and the
cosmic microwave background, as predicted by GZK
\cite{Greisen:1966jv}, or as products of the same astrophysical
sources that generate the cosmic rays observed in the highest
energy air showers \cite{Biermann:2002ja} -there is wide
expectation in the community that neutrinos of galactic or cosmic
origin will be detected in the UHE energy range in the near
future. Many experimental limits on astrophysical neutrino fluxes
have already been established.  Frejus,
%\cite{Rhode:1996es},
Baikal,
%\cite{Balkanov:2001ky},
AMANDA
% \cite{Ahrens:2003ee},
and MACRO \cite{Lowest E}
%\cite{Ambrosio:2002ma}
have reported limits on neutrinos from astrophysical sources in
the $10^{3}-10^{6}$ GeV range, while Fly's Eye,
%\cite{Baltrusaitis:1985mt},
AGASA,
% \cite{Yoshida:2001pw},
AMANDA
%\cite{Hundertmark:2003gf},
 and RICE \cite{Middle E}
 % \cite{Kravchenko:2003gj}
 have given limits in
the range $10^{6}-10^{9}$ GeV, and AGASA, Fly's Eye, RICE, Forte,
ANITA and GLUE \cite{highest E} have reported limits in the range
above $10^{9}$ GeV.  In the near future, IceCube \cite{icecube},
ANITA and AUGER \cite{auger} can be expected to release stronger
results based on more recent data. In the longer term, proposed
expansions of IceCube, such as AURA \cite{aura}, or a salt-based
radio telescope like SALSA \cite{salsa} could afford substantially
enhanced sensitivity. Further in the future, an orbiting telescope
like the proposed EUSO project \cite{euso} opens the possibility
to achieve a huge effective volume.

The design of neutrino telescopes depends critically on the
estimates of neutrino cross sections in the UHE regime, most of
which is far beyond currently available data. Moreover, the
systematic study of the scattering of such high energy neutrinos
with baryonic matter might prove to be an essential instrument to
test new physics effects, such as those given by models of low
scale gravity (LSG). Models of extra dimensions
\cite{Arkani-Hamed:1998rs,Randall:1999ee,Antoniadis:1998ig} have
lowered the characteristic quantum-gravity scale to energies
comparable with the electro-weak scale, enhancing the expected
neutrino cross sections at GZK-energies and above, even for
conservative parameter choices.

Efforts at realistic calculations of the UHE neutrino-nucleon
cross section have been performed in the framework of QCD and the
standard model \cite{Frichter:1994mx,Gandhi}, and many speculative
calculations based on new physics effects have been presented over
the past decade or more.  The only detailed calculations of
neutrino event rates within the framework of the LSG models that
include the neutrino-proton eikonal cross section with LSG
graviton exchange, rely upon the saddle point approximation
\cite{jain-hussain}.  In this article we present in detail the
calculation for the full eikonal approximation of the LSG
neutrino-proton cross section, including a study of the effects of
a finite tension on the brane \cite{bando,Sjodahl:2006gb}. We
comment on the inapplicability of the saddle point approximation
at neutrino energies below $10^{11}$ GeV.

In Sec. II we summarize the main procedure of the eikonal
approximation (EA) in quantum field theory and apply it to the
calculation of the desired differential cross sections at the
parton level in the Arkani-Hamed, Dimopoulous and Dvali (ADD)
model of extra dimensions \cite{Arkani-Hamed:1998rs}. We consider
the cases of an infinitely thin standard model brane and the one
where it has finite thickness. In Sec. III we present the full EA
calculation and compare the results with those in the saddle point
approximation, and with the standard model cross section. Sec. IV
is dedicated to the summary of results, comments on the
sensitivity of results to choice of gravity scale and brane
tension and on the origin of the discrepancy between the saddle
point and full EA calculation. We also comment on the impact of
our cross section results on neutrino telescope event rate
expectations. The Appendix presents details of the representation
of the eikonal amplitude in terms of Meijer G-functions.

\bigskip

\section{Theory}
%We will calculate the neutrino-nucleon cross section in the
%eikonal approximation for low scale gravity models.
The eikonal approximation is a technique widely used in quantum
mechanics and wave physics to derive an expression for elastic
scattering in the limit of large center of mass (CM) energies (see
for example Reference \cite{Sakurai}). In quantum field theory the
amplitude can be obtained by summation of all the ladder and
cross-ladder diagrams for boson exchange at all order
\cite{Cheng:1987ga,Kabat:1992tb}, as exemplified in Fig. 1.

\begin{figure}[hb!]
 \begin{center}
  \includegraphics[scale=1.5]{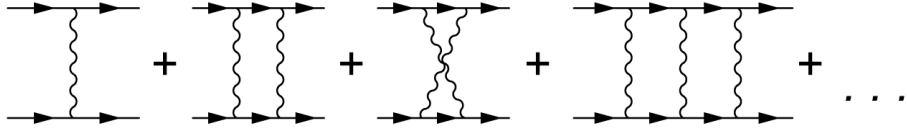}
     \caption{Sum of the ladder and cross-ladder diagrams in the eikonal approximation}
   \end{center}
\end{figure}

The approximation is valid for center of momentum energies that
are high with respect to $Q^2$, the absolute magnitude of the
momentum transfer $q^2$, i.e. $|t/s|\ll 1$ in the usual Mandelstam
variables for the $2\rightarrow 2$ scattering. The high energy
scattering angle is small and the four-momenta of the incoming
particles are approximately equal to those of the corresponding
outgoing particles, $p_{1}\sim p_{1}'$ and $p_{2}\sim p_{2}'$. For
any given kind of boson exchanged (scalar, vector, tensor), the
vertex and external legs factor is independent from the spin of
the incoming particles, as one can neglect their mass and the
recoil of the matter field.

With these assumptions, $s\sim 2(p_{1}\cdot p_{2})$ and the
tree-level amplitude (Fig. 2a), or Born term for the exchange of
one intermediate boson is given by
\begin{equation}
    A_{Born}(t)=g^{2}s^{r}\frac{1}{t-m^{2}},\label{gen_Born}
\end{equation}
where $g$ is the dimension-dependent interaction coupling
constant, $m$ is the mass of the exchanged particle and $r=0,1,2$
is its spin.

\begin{figure}[h!]
 \begin{center}
  \includegraphics[scale=0.4]{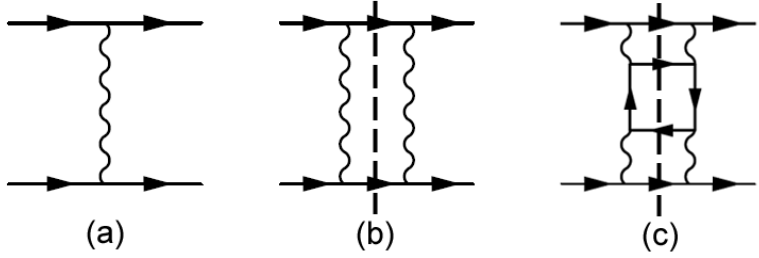}
     \caption{2a: Born term diagram. 2b: One loop diagram with matter propagators put on shell
     by the Cutkowsky rule. 2c: Example of an inelastic diagram.}
   \end{center}
\end{figure}

At one-loop level (Fig. 2b), the EA simplifies the matter
propagators:

\begin{equation}
    \frac{i}{(p\pm q')^{2}-M^{2}+i\varepsilon}\rightarrow\frac{i}{\pm2p\cdot
    q'+i\varepsilon},\label{matter_prop}
\end{equation}
where $M$ is the mass of the incoming particles. The one-loop
eikonal amplitude, obtained when the above propagators are put on
shell by the Cutkowsky rule, is imaginary.  The  integral over the
transferred four-momentum reduces to an integral over its
components in the plane perpendicular to the momenta of the
incoming particles, $q_{\perp}$. One gets for the eikonal at one
loop:

\begin{eqnarray}
    A_{loop}&=&-\frac{i}{2}(g^{2}s^{r})^{2}\int\frac{d^{4}q'}{(2\pi)^{4}}
    \frac{1}{q'^{2}-m^{2}}\frac{1}{(q-q')^{2}-m^{2}}(2\pi
    i)^{2}\delta(2p_{1}\cdot q')\delta(2p_{2}\cdot q')\nonumber\\
     &=&\frac{i}{4s}\int\frac{d^{2}q'_{\perp}}{(2\pi)^{2}}A_
 {Born}(q_{\perp}'^{2})A_{Born}[(q_{\perp}-q'_{\perp})^{2}]\nonumber\\
 &=&-2si\int d^{2}b_{\perp}e^{iq_{\perp}\cdot
b_{\perp}}\left(-\frac{1}{2}\chi^{2}\right)\label{one_loop}
\end{eqnarray}
where the eikonal phase
\begin{equation}
    \chi(b_{\perp})=\frac{1}{2s}\int\frac{d^{2}q_{\perp}'}{(2\pi)^{2}}e^{-iq_{\perp}'\cdot b_{\perp}}
    A_{Born}(q_{\perp}'^{2})\label{chi}
\end{equation}
is the Fourier transform of the Born term.

The procedure can be repeated with higher order diagrams
\cite{Cheng:1987ga}, such that one can sum all the terms in $\chi$
to obtain :

\begin{eqnarray}
    A_{eik}=A_{Born}+A_{loop}+\textrm{...}+A_{n-loops}+\textrm{...}&=&-2si\int
    d^{2}b_{\perp}e^{iq_{\perp}\cdot
    b_{\perp}}\left(\sum_{1}^{\infty}(i\chi)^{n}/n!\right)\nonumber\\
     &=&-2si\int d^{2}b_{\perp}e^{iq_{\perp}\cdot
    b_{\perp}}(e^{i\chi}-1).\label{eikonal}
\end{eqnarray}

It is worth noting at this point of our discussion that the
validity of the EA in describing the asymptotic behavior of
elastic scattering has been long debated. References
\cite{Cheng:1987ga} and \cite{Kabat:1992pz} argue that the eikonal
is not a correct description of the high energy limit in the cases
of scalar particle exchange, abelian vector exchange and
non-abelian vector exchange. In the cases of vector particle
exchange, when we put $r=1$ in Eq. (\ref{gen_Born}), the EA
describes accurately the asymptotic behavior of the sum of
exchange diagrams, but fails to consider the inelastic diagrams
related by unitarity to particle production, of which Fig. 2c is
an example. As has been shown by \cite{Cheng:1987ga}, such
diagrams become dominant at eighth order in $g$. On the other
hand, the $s^{2}$ dependence of the graviton-exchange Born term in
the quantum gravity EA renders the sum of exchange diagrams
dominant with respect to the inelastic unitarity diagrams
(\cite{Kabat:1992pz} and references therein). Consistent with the
state of present knowledge, we will consider the eikonal a good
approximation of the asymptotic behavior for graviton exchange.

In models of LSG with large extra dimensions
\cite{Arkani-Hamed:1998rs,Antoniadis:1998ig}, graviton exchange
($r=2$)
%- used as an input to the eikonal amplitude -
is ultraviolet divergent already at tree-level
\cite{Giudice:1998ck,Han:1998sg}. This is due to the presence of
an infinite tower of massive Kaluza-Klein modes over which the
amplitude has to be summed. For a number $n$ of extra-dimensions
we have, approximating the sum by an integral,

\begin{equation}
    A_{Born}=\frac{s^{2}}{M_{D}^{n+2}}\int\frac{d^{n}m}{t-m^{2}}
    \equiv \frac{s^{2}}{M_{D}^{n+2}}S_{n}\int_{0}^{\infty}\frac{m^{n-1}dm}
    {t-m^{2}},\label{lsg_Born}
\end{equation}
where $M_{D}$ is the $D$-dimensional Plank mass (where $D\equiv
4+n$) and $S_{n}$ is the surface of a $n$-dimensional unit sphere.
As the notation indicates, this sum is adopted as the "Born input"
to the eikonal calculation \cite{Emparan:2001ce}.

In the literature, there are two common solutions to the problem
of tree-level divergencies. One can introduce a cut-off of the
order of the low gravity scale, as this is the limit below which
the effective theory is deemed to be valid
\cite{Giudice:1998ck,Han:1998sg,Giudice:2001ce}. Or, one can
consider the physically reasonable case where the brane has a
finite tension $M_{S}$, corresponding to an extension $1/M_{S}$ of
the standard model fields along the extra dimensions
\cite{bando,Sjodahl:2006gb}. In this case, if the
extra-dimensional part $\vec{y}$ of the standard model wave
function has a gaussian cut-off of the kind

\begin{equation}
    \psi(\vec{y})=\left(\frac{M_{S}}{\sqrt{2\pi}}\right)^{\frac{n}{2}}
    e^{-\frac{|\vec{y}|^{2}M_{S}^{2}}{4}},\label{sm_wave_f}
\end{equation}
Eq. (\ref{lsg_Born}) is modified to

\begin{equation}
    A_{Born}=\frac{s^{2}}{M_{D}^{n+2}}S_{n}\int_{0}^{\infty}e^{-\frac{m^{2}}{M_{s}^{2}}}\frac{m^{n-1}dm}
    {t-m^{2}} .\label{Born_tens}
\end{equation}

The Born term in Eq. (\ref{lsg_Born}) can be calculated by
dimensional regularization \cite{Giudice:2001ce}:

\begin{equation}
    A_{Born}(q_{\perp}^{2})=\frac{\pi^{\frac{n}{2}}\Gamma\left(1-\frac{n}{2}\right)}{M_{D}^{n+2}}
    (q_{\perp}^{2})^{\frac{n}{2}-1}s^{2}.\label{dim_reg}
\end{equation}
Substitution of the latter into Eq. (\ref{chi}) yields

\begin{equation}
     \chi=\frac{\pi^{\frac{n}{2}-1}\Gamma(1-n/2)s}{4M_{D}^{n+2}}\int_{0}^{\infty}
    dqq^{n-1}J_{0}(qb)=\left[\frac{(4\pi)^{\frac{n}{2}-1}s\Gamma(\frac{n}{2})}{2M_{D}^{n+2}}\right]
    \frac{1}{b^{n}}\equiv\left(\frac{b_{c}}{b}\right)^{n}\label{grw_chi}
\end{equation}
where we've followed the commonly simplified notation
$q=q_{\perp}$, $b=b_{\perp}$. The critical impact parameter
$b_{c}$ separates the region of space where the Born term is
dominant ($b\gg b_{c}$) from the one where higher order terms in
$\chi$ become relevant. Expression (\ref{grw_chi}) is clearly
finite for all $b>0$. Thus, it can be seen from Eqs.
(\ref{lsg_Born})-(\ref{grw_chi}) that when the brane is infinitely
thin ($M_{S}\rightarrow\infty$) the ultraviolet divergence gets
absorbed by the eikonal phase calculation. Each regularization
procedure leads to the same finite result \cite{Giudice:2001ce}.
\newline
The eikonal amplitude is then given by
\begin{equation}
    A_{eik}=4\pi sb_{c}^{2}F_{n}(b_{c}q),\label{grw_eik}
\end{equation}
with
\begin{equation}
    F_{n}(\eta)=-i\int_{0}^{\infty}d\xi \xi
    J_{0}(\xi\eta)(e^{i\xi^{-n}}-1),\label{grw_F}
\end{equation}
and $J_{0}$ the Bessel function of order zero.\newline As
mentioned in passing in \cite{Giudice:2001ce,Emparan2}, but not
pursued, the $F_{n}$ can be expressed in terms of Meijer
G-functions. We implement our calculation explicitly in these
terms, providing details in the Appendix.  The result is
\begin{equation}
    F_{n}(\eta)=2^{-\frac{2}{n}-1}\pi^{\frac{1}{2}}n^{-1}(R_{n}(\eta)+iI_{n}(\eta)),\label{meijer}
\end{equation}
where the real and imaginary part are given by
%\begin{equation}
%    I_{n}(x)=G^{n+1,0}_{0,2(n+1)}\left(\frac{x^{2n}}{2^{2n+2}n^{2n}}\left|
%    \begin{array}{c}
%        0,1/n,...,(n-1)/n,(n-1)/n\\
%        -1/n,0,(n-2)/(2n),(n-2)/n,...,1/n
%    \end{array}
%        \right.\right)\label{im_meij}
%\end{equation}
%\begin{equation}
%    I_{n}(\eta)=G^{n+2,0}_{1,2n+3}\left(\frac{\eta^{2n}}{2^{2n+2}n^{2n}}\left|
%    \begin{array}{c}
%        \frac{3n-2}{2n}\\
%        -\frac{1}{n},\frac{1}{2}-\frac{1}{n},0,\frac{1}{n},...,\frac{n-1}{n},
%        0,\frac{3n-2}{2n},\frac{n-1}{n},...,\frac{1}{n}\end{array}
%    \right.\right)\label{im_meij}
%\end{equation}
\begin{equation}
    I_{n}(\eta)=G^{n+1,0}_{0,2(n+1)}\left(\frac{\eta^{2n}}{2^{2n+2}n^{2n}}\left|
    \begin{array}{c} \\
        0,\frac{1}{n},...,\frac{n-1}{n},\frac{n-1}{n},
        -\frac{1}{n},0,\frac{n-2}{2n},\frac{n-2}{n},...,\frac{1}{n}\end{array}
    \right.\right)\label{im_meij}
\end{equation}
and
%\begin{equation}
%    R_{n}(x)=G^{n+1,0}_{0,2(n+1)}\left(\frac{x^{2n}}{2^{2n+2}n^{2n}}\left|
%    \begin{array}{c}
%        0,(n-2)/(2n),1/n,...,(n-1)/n\\
%        -1/n,0,(n-1)/n,(n-2)/n,...,1/n
%    \end{array}
%    \right.\right).\label{re_meij}
%\end{equation}\bigskip
%\begin{equation}
%    R_{n}(\eta)=G^{n+2,0}_{1,2n+3}\left(\frac{\eta^{2n}}{2^{2n+2}n^{2n}}\left|
%    \begin{array}{c}
%        \frac{n-1}{n}\\
%        -\frac{1}{n},\frac{1}{2}-\frac{1}{n},0,\frac{1}{n},...,\frac{n-1}{n},
%        0,\frac{n-1}{n},\frac{n-1}{n},...,\frac{1}{n}\end{array}
%    \right.\right),\label{re_meij}
%\end{equation}
\begin{equation}
    R_{n}(\eta)=-G^{n+1,0}_{0,2(n+1)}\left(\frac{\eta^{2n}}{2^{2n+2}n^{2n}}\left|
    \begin{array}{c} \\
        0,\frac{n-2}{2n},\frac{1}{n},...,\frac{n-1}{n},
        -\frac{1}{n},0,\frac{n-1}{n},\frac{n-2}{n},...,\frac{1}{n}\end{array}
    \right.\right),\label{re_meij}
\end{equation}
where $G_{ij}^{kl}$ are the Meijer G-functions \cite{Bateman}.

In the limit of large momentum transfer with respect to the
inverse impact parameter ($q\gg 1/b_{c}$), the amplitude of Eq.
(\ref{grw_eik}) can be approximated analytically by the steepest
descent method
(\cite{jain-hussain,Emparan:2001ce,Giudice:2001ce,Emparan2}) and
it reads
\begin{equation}
A_{eik}(q)=A_{n}e^{i\phi_{n}}\left[\frac{s}{qM_{D}}\right]^{\frac{n+2}{n+1}},\label{sad_point}
\end{equation}
where
\begin{equation}
A_{n}=\frac{(4\pi)^{\frac{3n}{2n+2}}}{\sqrt{n+1}}
\left[\Gamma\left(\frac{n}{2}+1\right)\right]^{\frac{1}{n+1}},\label{spAmpl}
\end{equation}
\begin{equation}
\phi_{n}=\frac{\pi}{2}+(n+1)\left[\frac{qb_{c}}{n}\right]^{\frac{n}{n+1}},\label{spPhase}
\end{equation}
and $b_{c}$ is given by Eq. (\ref{grw_chi}).  The location of the
saddle point, $b_{s}$ =$b_{c}$
$(\frac{n}{qb_{c}})^{\frac{1}{n+1}}$, should satisfy $b_{s} \ll
b_{c}$ in order for the saddle point to give the dominant
contribution to the amplitude. \bigskip

On the other hand, in the case of a finite tension $M_{S}$ in the
brane, substitution of Eq. (\ref{Born_tens}) into Eq. (\ref{chi})
yields \cite{Sjodahl:2006gb}:
\begin{equation}
    \chi(b)=\frac{sM_{S}^{n}}{M_{D}^{n+2}}\Gamma\left(\frac{n}{2}\right)\frac{\pi^{\frac{n}{2}-1}}{8}
    U\left(\frac{n}{2},1;\frac{M_{S}^{2}b^{2}}{4}\right)
    =\left(\frac{b_{c}M_{S}}{2}\right)^{n}U\left(\frac{n}{2},1;\frac{M_{S}^{2}b^{2}}{4}\right),\label{sjod_chi}
\end{equation}
where $U(a,c;x)$ is a confluent hypergeometric function of the
second kind \cite{Bateman}.

By substitution of Eq. (\ref{sjod_chi}) into Eq. (\ref{eikonal})
the eikonal amplitude can be calculated easily:
\begin{equation}
    A_{eik}=-4\pi is\int_{0}^{\infty}db b J_{0}(qb)\left[e^{i
    \left(\frac{b_{c}M_{S}}{2}\right)^{n}U\left(\frac{n}{2},1;\frac{M_{S}^{2}b^{2}}{4}\right)}-1\right],\label{sjod_eik}
\end{equation}
where in the last expression we have used the functional identity:
\begin{equation}
    \int d^{2}\eta e^{i \vec{\xi}\cdot\vec{\eta}}f(\eta)=2\pi
    \int_{0}^{\infty}d\eta \eta J_{0}(\xi \eta)f(\eta),
\end{equation}
with $\xi=|\vec{\xi}|$, $\eta=|\vec{\eta}|$. We note here that, to
our knowledge, there is no saddle point approximation for the
extended brane case in the literature. Nor we were able to find
for Eq. (\ref{sjod_eik}) a simple, useful closed-form like
(\ref{sad_point})-(\ref{spPhase}), as the solution requires the
numerical treatment of a transcendental equation.

The differential cross section at the parton level can now be
calculated with the usual substitution $-t\equiv -q^{2}\equiv
Q^{2}=xys$, where $x$ is the fraction of momentum of the proton
carried by the parton and $y$ is the inelasticity. We get for the
infinitely thin brane:
\begin{equation}
    \frac{d^{2}\sigma}{dxdy}=\sum_{i}xf_{i}(x, xys)s\pi
    b_{c}^{4}(\hat{s})|F_{n}(b_{c}(\hat{s})\sqrt{xys})|^{2},\label{grw_xsect}
\end{equation}
with $F_{n}(\eta)$ given by Eq. (\ref{grw_F}).\newline For the
brane $1/M_{S}$-thick:
\begin{equation}
    \frac{d^{2}\sigma}{dxdy}=\sum_{i}xf_{i}(x,
    xys)s\pi|G_{n}(\sqrt{xys})|^{2},\label{sjod_xsect}
\end{equation}
with
\begin{equation}
    G_{n}(\zeta)=-i\int_{0}^{\infty}db b J_{0}(b\zeta)\left[e^{i
    \left(\frac{b_{c}(\hat{s})M_{S}}{2}\right)^{n}U\left(\frac{n}{2},1,\frac{M_{S}^{2}b^{2}}{4}\right)}-1\right].\label{sjod_G}
\end{equation}
The $f_{i}(x,Q^{2})$ are the usual parton distribution functions
(PDF) summed over all the quark and antiquark flavors and the
gluons. Note here that the parameter $b_{c}(s,n)$ of Eq.
(\ref{grw_chi}) has to be calculated with respect to the center of
mass energy squared of the neutrino-parton system $\hat{s}=xs$.

\section{Results and Discussion}

The total cross section is obtained by integration of Eqs.
(\ref{grw_xsect}) and (\ref{sjod_xsect}). The $y$-integral has
been performed with Mathematica using the CTEQ5 PDFs
\cite{Lai:1999wy}, in the interval
$[Q_{0}^{2}/(xs),1/(xsR_{S}^{2})]$, where
$Q_{0}^{2}=0.01M_{W}^{2}$, $M_{W}$ is the mass of the $W$-boson
and $R_{S}^{2}\propto s^{1/(n+1)}/M_{D}^{2(n+2)/(n+1)}$ is the
Schwarzschild radius for black-hole production in the
$n$-dimensional theory \cite{Giudice:2001ce}. We adopt
$\sqrt{Q^{2}}$ as the scale parameter in the PDF's. The upper
limit of integration is introduced since the eikonal cross section
is not applicable for $Q^{2}\gtrsim1/R_{S}^{2}$. The lower limit,
as in standard model calculations, allows us to be safely inside
the perturbative QDC regime. The $x$-integral has been performed
by dividing the interval $[Q_{0}^{2}/s,1]$ into 38 logarithmic
subintervals and performing Simpson's rule. This procedure gives
an accurate evaluation of the cross section, which is reasonably
smooth in $x$.\bigskip

We consider the case of $n=5$ extra-dimensions, which is
sufficient to illustrate all our points, since by changing $n$ the
calculations differ by amounts small compared to changes due to
$M_{D}$, $M_{S}$. In Fig. 3 we show the comparison between our
calculation of the proton-neutrino cross section in the full EA
with the saddle point approximation of the same cross-section
given by Eqs. (\ref{sad_point}), (\ref{spAmpl}) and
(\ref{spPhase})\footnote{For an appropriate comparison with the
calculations of Ref. \cite{jain-hussain}, when evaluating the
total cross section we replaced the eikonal amplitude in Eq.
(\ref{grw_xsect}) by the Born term for $x\leq M_{D}^{2}/s$. For a
detailed treatment of the Born term, see the first paper of Ref.
\cite{jain-hussain}.}. The brane is here assumed to be infinitely
thin, with the (4+5)-dimensional Planck scale set at $M_{D}=1$
TeV. In the range $E_{\nu}>10^{7}$ GeV, where the EA is expected
to be valid, the full eikonal cross section is significantly
larger than its evaluation at the saddle point. The difference
ranges from a factor 2.6 at the low end to about 1.5 at the high
end. As discussed in the next section, the conditions for the
dominance of the saddle point evaluation are not satisfied until
extremely high energies, explaining the slow convergence of the
saddle point value to the full eikonal result. We point out here
that the calculation of the saddle point approximation at $n=6$
differs from the one at $n=5$ by $10-20\%$ over the entire energy
range, in accordance with the findings of Ref.
\cite{jain-hussain}. As it will be even clearer in what follows,
this is a small amount in comparison to how the other parameters
affect the results.
\begin{figure}[ht!]
 \begin{center}
  \includegraphics[width=80mm, height=60mm]{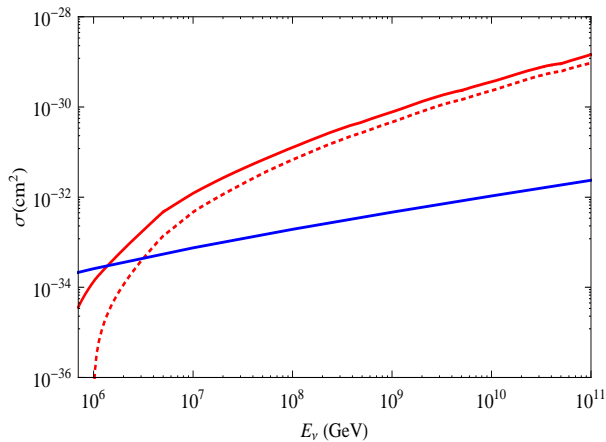}
     \caption{Solid light line: Full EA corrected by the Born term$^{1}$, $M_{D}=10^{3}$ GeV, $M_{S}=\infty$, $n=5$.
     Dashed light line: Saddle point approximation corrected by the Born term. Solid dark line: Standard model neutral current.}
   \end{center}
\end{figure}

As also shown in Fig. 3, even with the saddle point approximation
to the LSG eikonal, the neutral current cross section in this
model exceeds the standard model values \cite{Gandhi} at CM
energies above the gravity scale, corresponding to roughly $10^6$
GeV for $M_{D}$ = 1 TeV \cite{jain-hussain}. The dominance is even
more striking when the full eikonal is used: at neutrino energies
greater than $10^{10}$ GeV new physics effects raise the cross
section by almost three orders of magnitude.
%\begin{figure}[ht!]
% \begin{center}
%  \includegraphics[width=80mm, height=60mm]{FinEik+nc_c.eps}
%     \caption{Light line: Full EA corrected by the Born term, $M_{D}=10^{3}$ GeV, $M_{S}=\infty$, $n=5$.
%     Dark line: standard model neutral current.}
%   \end{center}
%\end{figure}

The presence of a thick brane with a tension $M_{S}=1$ TeV sharply
reduces the cross section, due to the effective gaussian cut-off
in the Born amplitude of Eq. (\ref{Born_tens}). The lower tension
suppresses the eikonal cross section by an order of magnitude at
the low energy end and a factor of two at the high end, as shown
in Fig. 4. As the tension rises, and the brane gets thinner, the
cross section should approach the infinitely thin limit. We
already obtain an almost perfect overlapping when the tension is
$M_{S}=10$ TeV.
\begin{figure}[ht!]
 \begin{center}
  \includegraphics[width=80mm, height=60mm]{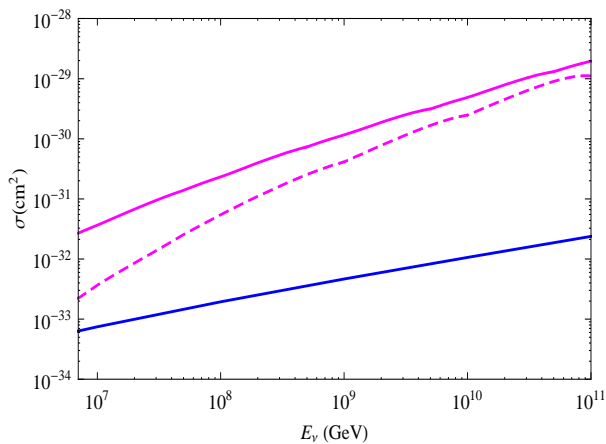}
     \caption{Eikonal cross sections at LSG $M_{D}=10^{3}$ GeV for branes with different tension. (Not corrected by the Born
     term.)
     Solid light: $M_{S}=\infty$, $n=5$.
     Dashed light: $M_{S}=10^{3}$ GeV, $n=5$. Solid dark: Standard model NC.}
   \end{center}
\end{figure}
%This are the new plots:
%\begin{figure}[ht!]
% \begin{center}
%  \includegraphics[width=80mm, height=60mm]{Fincomp10TeVGRWb.eps}
%     \caption{Eikonal cross sections at LSG $M_{D}=10^{3}$ GeV for branes with different
%    tension: $M_{S}=\infty$ and $M_{S}=10^{4}$ GeV, $n=5$.}
%   \end{center}
%\end{figure}

Fig. 5 shows the reduction to the cross sections due to a higher
Planck scale $M_{D}=2$ TeV. We here display the cases of the
infinitely thin brane and of the $M_{S}=1$ TeV-tension brane. We
see that the increase in scale to 2 TeV suppresses the cross
section an order of magnitude over the entire energy range for the
thin brane case (solid light line). The relative suppression is
even bigger in the case of a thick brane (dashed light line),
where the cross section is lowered to values beneath the standard
model calculation at energies less than $10^{9}$ GeV.
\begin{figure}[h!]
 \begin{center}
  \includegraphics[width=80mm, height=60mm]{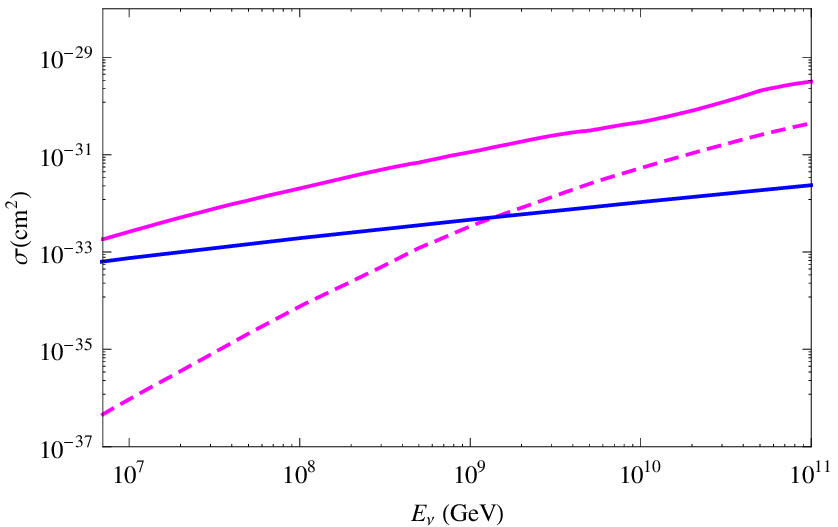}
     \caption{Eikonal cross sections at LSG $M_{D}=2\times10^{3}$ GeV for branes with different tension. (Not corrected by the Born
     term.)
     Solid light: $M_{S}=\infty$, $n=5$.
     Dashed light: $M_{S}=10^{3}$ GeV, $n=5$. Solid dark: Standard model NC.}
   \end{center}
\end{figure}

%\begin{figure}[h!]
% \begin{center}
%  \includegraphics[width=80mm, height=60mm]{M1M2nc_b.eps}
%     \caption{Pure EA comparison at high energies for different LSG and equal tension.
%     Solid light: $M_{D}=10^{3}$ GeV, $M_{S}=\infty$, $n=5$.
%     Dashed light: $M_{D}=2\times10^{3}$, $M_{S}=\infty$ GeV, $n=5$. Solid dark: Standard model NC.}
%   \end{center}
%\end{figure}
%\begin{figure}[h!]
% \begin{center}
%  \includegraphics[width=80mm, height=60mm]{TensM1M2nc_b.eps}
%     \caption{Pure EA comparison at high energies for different LSG and equal tension.
%     Solid light: $M_{D}=10^{3}$ GeV, $M_{S}=10^{3}$ GeV, $n=5$.
%     Dashed light: $M_{D}=2\times10^{3}$ GeV, $M_{S}=10^{3}$ GeV, $n=5$.
%     Solid dark: Standard model NC.}
%   \end{center}
%\end{figure}

\section{Summary, Further Discussion and Conclusions}
We have numerically calculated the parton level cross-sections for
UHE neutrino-nucleon gravitational scattering in the eikonal
approximation, under the assumption that the standard model
particles are confined in a 4-dimensional brane embedded in a
higher number of compactified extra-dimensions (ADD Model). The
analytical calculations are presented in the most general case of
$n$ extra-dimensions, but the numerical calculations were
performed for $n=5$ to illustrate the magnitude of the effects for
a value of the extra dimensions that is only weakly constrained by
current laboratory and astrophysical data ($n\geq 5$). Lower
dimensions are required by astrophysical data to have
corresponding gravity scales too high to allow significant effects
on the UHE cross sections, whereas $n\geq 5$ are constrained only
by collider data, which put lower limits on the gravity scale less
than a TeV \cite{hannestad,D0}. The cross sections are only mildly
dependent on $n$, at any rate \cite{jain-hussain}. Previous
applications of the ADD-type models to the UHE neutrino cross
section and expectations for event rates relied on the saddle
point estimation of the eikonal amplitude, and only in the thin
brane limit. Our application completes the picture of the impact
of low scale gravity on the predictions of high energy
neutrino-nucleon cross sections.  We reach the same qualitative
conclusions as previous studies: the effects are sensitive to the
scale, or radius, of the extra dimensions but much less sensitive
to the actual number of dimensions. We also find that the brane
tension, or thickness, greatly affects the size of LSG influence
on UHE neutrino scattering, weakening sharply as brane tension
decreases, or equivalently as brane thickness increases. This is
in agreement with similar predictions for hadron collider
applications \cite{Sjodahl:2006gb}.

Quantitatively speaking, we have found that the presence of new
physics at LSG $M_{D}=10^{3}$ GeV enhances the "neutral current"
cross section compared to the standard model cross section by
three order of magnitude at CM energies of $10^{10}$ GeV and
above, and that the full eikonal calculation of the cross section
gives values consistently higher than the corresponding saddle
point approximation, when the brane is considered to be infinitely
thin. The enhancement above the standard model expectation is not
nearly as marked in the case of a brane physically extended
$10^{-3}$ GeV$^{-1}$ along the extra-dimensions, or in the case
where the LSG Planck Mass is increased by a factor of two.
However, we found that calculations of the cross sections
performed with brane tensions of 10 TeV and above are
indistinguishable, indicating that a ratio of tension to gravity
scale of 10:1 is effectively "infinite" for these purposes.

The preceding statement is quite $n$-independent as can be seen
from the r.h.s. of Eq. (\ref{sjod_chi}): since in the region of
validity of the EA ($s>M_{D}^2$) the impact parameter $b$ is of
the order of $b_{c}$, one can see that when $M_{S}\gg b_{c}^{-1}$
the hypergeometric function $U$ tends rapidly to
$(M_{S}b_{c}/2)^{-n}$. The eikonal phase is therefore of order
one, as expected, independently from the number of
extra-dimensions. Correspondingly, the approach to the thin brane
limit is only weakly dependent on $n$.

To understand the difference between the UHE cross section
evaluated in the saddle point approximation versus the full
eikonal calculation, we recall that the saddle point dominance of
the integral in question requires $q b_{c} \gg 1$.  In our eikonal
evaluation, as in those in Ref. \cite{jain-hussain}, we required
$q \leq 1/R_{S}$.  Let us check the consistency of these two
requirements.  Defining $q_{max}=1/R_{S}$, we find that $0.77 \leq
q_{max} \times b_{c} \leq 5.1$ as $E_{\nu}$ runs from $10^5$ GeV
to $10^{12}$ GeV for $M_{D}$ = $10^3$ GeV and $n=5$. The ranges
are $0.71 \leq q_{max} \times b_{c} \leq 8.0$ and $0.81 \leq
q_{max} \times b_{c} \leq 3.8$ for $n=4$ and $n=6$ respectively.
Moreover, the saddle point value of the impact parameter,
$b_{s}=b_{c}(qb_{c}/n)^{-1/(n+1)}$ is only slightly less than
$b_{c}$ for these parameter choices, failing to satisfy the
condition $b_{s}\ll b_{c}$ needed for saddle point dominance of
the phase integral, Eq. (\ref{grw_F}). These considerations tell
us that, even with the maximal $q$ consistent with use of the
eikonal, the strong coupling, saddle point conditions are, at
best, only marginally satisfied even at the highest, super GZK
energies we consider.

The sensitivity of the calculation to $M_{S}$ and $M_{D}$ values
is much more important to the estimate of event rates than the
sensitivity to the use of the saddle point approximation, rather
than the full EA. Since collisions that satisfy the EA conditions
tend to be highly elastic, the consequences for event rate
estimates based on the model assumptions are not severe, since the
fraction of deposited energy is small, and the corresponding
events are mixed with highly inelastic events at neutrino energies
an order of magnitude or more lower, where the flux is presumably
much larger. Thus the factor of ten difference in cross section
between the $M_{D}=$ 1 TeV and 2 TeV, or between the infinite and
finite tension at fixed LSG is, in a sense, part of uncertainty in
a "background" correction to inelastic processes, such as black
hole formation, where the neutrino energy is nearly all converted
to visible energy. Only if black hole formation is severely
suppressed, does the enhanced signal from elastic processes become
a sensitive tool to detect LSG.

In summary, our conclusions are that, in agreement with other
studies in collider settings, the LSG effects are sensitively
dependent on the choice of scale parameters - the gravity scale
and the brane tension. The eikonal neutrino-nucleon cross section
always rises sharply above the standard model versions, but the
energy at which the onset occurs and the degree to which the new
physics dominates, is strictly dictated by the input scales.  In
the regime we consider, $10^6 - 10^{12}$ GeV, the saddle point
dominance conditions are not well satisfied, and the
approximation, though lending itself to simple expressions and
intuitive pictures, noticeably underestimates the cross section.
Having said that, the uncertainty in $M_{D}$ and $M_{S}$ values is
nevertheless the dominant effect.

{\em Acknowledgements:} We thank Danny Marfatia for many helpful
conversations on this work. E. S. is supported in part by NSF
Grant No. PHY-0544278. D.W.M. receives support from DOE Grant No.
DE-FG02-04ER41308.

\section*{Appendix}
The integral of Eq. (\ref{grw_F}) can be expressed as a Mellin
convolution:
\begin{equation}
    F_{n}(y)=i\int_{0}^{\infty}\frac{dx}{x}f_{1}(x)f_{2}\left(\frac{y}{x}\right),\label{mellin}
\end{equation}
where
\begin{eqnarray}
    f_{1}(x)&=&\frac{1}{x^{2}}\left(e^{ix^{n}}-1\right)\label{f1}\\
    f_{2}(x)&=&J_{0}\left(x\right).\label{f2}
\end{eqnarray}

We can calculate the Mellin transforms of Eqs. (\ref{f1}),
(\ref{f2}):
\begin{equation}
    f_{1}^{\ast}(s)=\int_{0}^{\infty}f_{1}(x)x^{s-1}dx=
    -\frac{1}{n}\Gamma\left(\frac{s-2}{n}\right)e^{i\frac{\pi}{n}\left(n-1+\frac{s}{2}\right)}
    \label{trans1}
\end{equation}
\begin{equation}
    f_{2}^{\ast}(s)=\int_{0}^{\infty}f_{2}(x)x^{s-1}dx=
    \frac{2^{s-1}\Gamma\left(\frac{s}{2}\right)}{\Gamma\left(1-\frac{s}{2}\right)}.
    \label{trans2}
\end{equation}
Upon defining $f^{\ast}(s)=f_{1}^{\ast}(s)f_{2}^{\ast}(s)$, the
integral of Eq. (\ref{mellin}) turns out to be a
\textit{Mellin-Barnes} integral:
\begin{equation}
    -iF_{n}(y)=\frac{1}{2\pi i}\oint f^{\ast}(s)y^{-s}ds\label{mellin_barnes}
\end{equation}
that can be expressed explicitly in terms of Meijer G-functions.

 The real part of Eq. (\ref{mellin_barnes}) is
\begin{eqnarray}
    \Re F_{n}(y)&=&-\frac{1}{2\pi i}\oint\Im
    f^{\ast}(s)y^{-s}ds\nonumber\\
     &=&-\frac{1}{2\pi i}\oint-\frac{1}{n}
    \frac{2^{s-1}\Gamma\left(\frac{s}{2}\right)\Gamma
    \left(\frac{s-2}{n}\right)\sin\left[\frac{\pi}{n}\left(n-1+\frac{s}{2}\right)\right]}
    {\Gamma\left(1-\frac{s}{2}\right)}y^{-s}ds\nonumber\\
     &=&\frac{1}{2\pi i}\oint\frac{2^{s-1}\pi\Gamma\left(\frac{s}{2}\right)\Gamma
    \left(\frac{s-2}{n}\right)}
    {n\Gamma\left(1-\frac{s}{2}\right)\Gamma\left(\frac{n-1}{n}+\frac{s}{2n}\right)
    \Gamma\left(1-\frac{n-1}{n}-\frac{s}{2n}\right)}y^{-s}ds,\label{ReFstep1}
\end{eqnarray}
where in the last step we have used Euler's duplication formula
($\sin(\pi x)=\pi/[\Gamma(x)\Gamma(1-x)]$).

In the standard definition of Meijer G-functions \cite{Bateman}
the coefficient multiplying $s$ inside the Gamma functions has to
be $\pm1$. We therefore perform the substitution $s\rightarrow2ns$
to get
\begin{equation}
    \Re F_{n}(y)=\frac{1}{2\pi i}\oint\frac{\pi\Gamma(ns)\Gamma\left(2\left(s-\frac{1}{n}\right)\right)}
    {\Gamma(1-ns)\Gamma\left(\frac{n-1}{n}+s\right)\Gamma\left(1-\frac{n-1}{n}-s\right)}
    \left[\left(\frac{y}{2}\right)^{2n}\right]^{-s}ds,\label{ReFstep2}
\end{equation}
and use some known identities of the Gamma function to recast Eq.
(\ref{ReFstep2}) into
%\begin{multline}
%    \Re F_{n}(y)=2^{-\frac{2}{n}-1}\pi^{\frac{1}{2}}n^{-1}\textrm{ }\times\\
%    \frac{1}{2\pi i}\oint\frac{\Gamma\left(s-\frac{1}{n}\right)\Gamma\left(s-\frac{1}{n}+\frac{1}{2}\right)
%    \Gamma(s)\Gamma\left(s+\frac{1}{n}\right)...\Gamma\left(s+\frac{n-1}{n}\right)}
%    {\Gamma(1-s)\Gamma\left(\frac{1}{n}-s\right)...\Gamma\left(\frac{n-1}{n}-s\right)
%    \Gamma\left(\frac{n-1}{n}+s\right)\Gamma\left(1-\frac{n-1}{n}-s\right)}
%    \left(\frac{y^{2n}}{2^{2n+2}n^{2n}}\right)^{-s}ds\label{ReFstep3}
%\end{multline}
%The integral is now in the form of a Meijer G-function:
%\begin{multline}
%    \Re F_{n}(y)=2^{-\frac{2}{n}-1}\pi^{\frac{1}{2}}n^{-1}\textrm{ }\times\\
%    G^{n+2,0}_{1,2n+3}\left(\frac{y^{2n}}{2^{2n+2}n^{2n}}\left|
%    \begin{array}{c}
%        \frac{n-1}{n}\\
%        -\frac{1}{n},\frac{1}{2}-\frac{1}{n},0,\frac{1}{n},...,\frac{n-1}{n},
%        0,\frac{n-1}{n},\frac{n-1}{n},...,\frac{1}{n}\end{array}
%    \right.\right).\label{finalReF}
%\end{multline}
%A similar calculation for the imaginary part will yield:
%\begin{multline}
%    \Im F_{n}(y)=2^{-\frac{2}{n}-1}\pi^{\frac{1}{2}}n^{-1}\textrm{ }\times\\
%    G^{n+2,0}_{1,2n+3}\left(\frac{y^{2n}}{2^{2n+2}n^{2n}}\left|
%    \begin{array}{c}
%        \frac{3n-2}{2n}\\
%        -\frac{1}{n},\frac{1}{2}-\frac{1}{n},0,\frac{1}{n},...,\frac{n-1}{n},
%        0,\frac{3n-2}{2n},\frac{n-1}{n},...,\frac{1}{n}\end{array}
%    \right.\right).\label{finalImF}
%\end{multline}
\begin{multline}
    \Re F_{n}(y)=-2^{-\frac{2}{n}-1}\pi^{\frac{1}{2}}n^{-1}\textrm{ }\times\\
    \frac{1}{2\pi i}\oint\frac{\Gamma\left(s-\frac{1}{n}+\frac{1}{2}\right)
    \Gamma(s)\Gamma\left(s+\frac{1}{n}\right)\Gamma\left(s+\frac{2}{n}\right)...
    \Gamma\left(s+\frac{n-1}{n}\right)}{\Gamma(1-s)\Gamma\left(\frac{1}{n}-s\right)
    \Gamma\left(\frac{2}{n}-s\right)...\Gamma\left(\frac{n-1}{n}-s\right)
    \Gamma\left(1+\frac{1}{n}-s\right)}
    \left(\frac{y^{2n}}{2^{2n+2}n^{2n}}\right)^{-s}ds.\label{ReFstep3}
\end{multline}
The integral is now in the form of a Meijer G-function:
\begin{multline}
    \Re F_{n}(y)=-2^{-\frac{2}{n}-1}\pi^{\frac{1}{2}}n^{-1}\textrm{ }\times\\
    \textrm{MeijerG}\left[\left\{\left\{\right\},\left\{\right\}\right\},
    \left\{\left\{0,\frac{n-2}{2n},\frac{1}{n},...,\frac{n-1}{n}\right\},
    \left\{-\frac{1}{n},0,\frac{n-1}{n},\frac{n-2}{n},...,\frac{1}{n}\right\}\right\},
    \frac{y^{2n}}{2^{2n+2}n^{2n}}\right]\equiv\\
    -2^{-\frac{2}{n}-1}\pi^{\frac{1}{2}}n^{-1}\textrm{ }\times
    G^{n+1,0}_{0,2(n+1)}\left(\frac{y^{2n}}{2^{2n+2}n^{2n}}\left|
    \begin{array}{c} \\
        0,\frac{n-2}{2n},\frac{1}{n},...,\frac{n-1}{n},-\frac{1}{n},0,\frac{n-1}{n},\frac{n-2}{n},...,\frac{1}{n}
        \end{array}
    \right.\right).
    \label{finalReF}
\end{multline}
A similar calculation for the imaginary part yields
\begin{multline}
    \Im F_{n}(y)=2^{-\frac{2}{n}-1}\pi^{\frac{1}{2}}n^{-1}\textrm{ }\times\\
    \textrm{MeijerG}\left[\left\{\left\{\right\},\left\{\right\}\right\},
    \left\{\left\{0,\frac{1}{n},...,\frac{n-1}{n},\frac{n-1}{n}\right\},
    \left\{-\frac{1}{n},0,\frac{n-2}{2n},\frac{n-2}{n},...,\frac{1}{n}\right\}\right\},
    \frac{y^{2n}}{2^{2n+2}n^{2n}}\right]\equiv\\
    2^{-\frac{2}{n}-1}\pi^{\frac{1}{2}}n^{-1}\textrm{ }\times
    G^{n+1,0}_{0,2(n+1)}\left(\frac{y^{2n}}{2^{2n+2}n^{2n}}\left|
    \begin{array}{c} \\
        0,\frac{1}{n},...,\frac{n-1}{n},\frac{n-1}{n},-\frac{1}{n},0,\frac{n-2}{2n},\frac{n-2}{n},...,\frac{1}{n}
        \end{array}
    \right.\right).\label{finalImF}
\end{multline}

In Eqs. (\ref{finalReF}) and (\ref{finalImF}), the first line
refers to the Mathematica input necessary to perform the
calculation, while the second line refers to the standard
mathematical notation of Eqs. (\ref{im_meij}) and (\ref{re_meij})
\cite{Bateman}.

\end{document}